\input{psfig.sty}
\input{epsf.sty}
\documentclass[twocolumn,showpacs,preprintnumbers,amsmath,floatfix]{revtex4}

\usepackage{graphicx}
\usepackage{dcolumn}
\usepackage{bm}

\begin{document}

\title{Quark mass dependence of the pseudoscalar hairpin vertex}
\author{W.~A.~Bardeen$^{a}$, E.~Eichten$^{a}$, and H.B.~Thacker$^{b}$}
\affiliation{
 $^{a}$ Fermi National Accelerator Laboratory, P.O.~Box 500, Batavia, IL 60510 \\
 $^{b}$Department of Physics, University of Virginia, Charlottesville, VA 22901 }
\date{May 20, 2004}


\begin{abstract}
In a recent investigation of chiral behavior in quenched lattice QCD, the flavor-singlet 
pseudoscalar ``hairpin'' vertex associated with the $\eta'$ mass was studied for pion masses 
ranging from $\approx 275$ to $675$ MeV. Throughout this mass range, 
the quark-disconnected pseudoscalar correlator is well-described by a pure double-pion-pole diagram with
a $p^2$-independent mass insertion. The
residue of the double pole was found to exhibit a significant quark mass dependence, evidenced by
a negative slope of the effective mass insertion $(m_0^{eff})^2$ as a function of $m_{\pi}^2$.
It has been observed by Sharpe that, with a consistent NLO calculation in
quenched chiral perturbation theory, this
mass dependence is uniquely predicted in terms of the single-pole coefficient $\alpha_{\Phi}$ and the
Leutwyler parameter $L_5$. Since $\alpha_{\Phi}$ is found to be $\approx 0$, the chiral slope of the
double-pole residue determines a value for $L_5$. 
This provides a consistency check between the chiral slope of the
hairpin mass insertion and that of the pion decay constant. We investigate the consistency of these 
mass dependences in our Monte Carlo results at
two values of lattice spacing. Within statistics, the slopes are found to be consistent with the
$Q\chi PT$ prediction, confirming that the observed negative slope of $m_0^{eff}$ arises as
an effect of the $L_5$ Leutwyler term. 
\end{abstract}
\maketitle


The predictions of chiral perturbation theory provide extensive tests of numerical calculations for light-quark 
amplitudes in lattice QCD. In a chiral Lagrangian framework, the quenched approximation introduces a variety of anomalous effects  
associated with the flavor singlet pseudoscalar $\eta'$ meson. These quenched chiral loop effects
can serve as a useful diagnostic for lattice QCD calculations with light quarks. Successfully reproducing the quark
mass dependences predicted by chiral perturbation theory provides an important test of the practical validity
of any chiral lattice fermion method. Although the theoretical underpinnings of quenched chiral perturbation theory
($Q\chi PT$) \cite{B&G,Sharpe} are less rigorous than for full QCD, recent lattice calculations \cite{BDET1,BDET2,BET} have exhibited
good agreement with $Q\chi PT$ predictions in a number of amplitudes involving scalar and pseudoscalar meson correlators. 
Within the mass range studied ($m_{\pi}\approx 275$ to $675$ MeV), these results provide strong support for the assertion that 
the effect of the chiral $U(1)$ anomaly is well-represented by a $p^2$-independent
$\eta'$ mass insertion, which can be treated perturbatively according to the
rules of $Q\chi PT$. 

A direct study of the time-dependence of the singlet pseudoscalar correlator shows that 
the hairpin insertion vertex is a pure $\eta'$ mass term, $\propto m_0^2\eta'^2$, with 
higher derivative terms being suppressed. The $\alpha_{\Phi}$ parameter of a 
possible $\alpha_{\Phi}(\partial_{\mu}\eta')^2$ term is found to be 
consistent with $0$ and definitely $<<1$. 
Specifically, the correlator fits well at all time separations to a pure double-pion pole formula, 
with little evidence for either a single pole term (parametrized by $\alpha_{\Phi}$) or for any excited-state contamination.
An interesting feature of the results presented in \cite{BET} is that the effective $\eta'$ mass insertion $(m_0^{eff})^2$ obtained from the
residue of the double pole decreases significantly with increasing quark mass. The possibility of explaining this mass dependence
as an effect of the $p^2$-dependent $\alpha_{\Phi}$ term is ruled out by the shape of the time dependence of the 
hairpin correlator, which gives $\alpha_{\Phi}\approx 0$.

A consistent explanation of the mass dependence of the double-pole
residue has recently been proposed by Sharpe \cite{Sharpe_hp}, who observed that the value of $(m_0^{eff})^2$ receives an order
$m_{\pi}^2$ contribution from the $L_5$ Leutwyler term. Sharpe also showed that
any other possible terms of order $m_q$ or $p^2$ that could contribute to the mass 
dependence  of the hairpin double pole residue can be removed by a field 
redefinition along with a redefinition of $\alpha_{\Phi}$. 
As a result, such terms do not introduce any new physical parameters, and therefore the lowest order quark mass 
dependence of the hairpin insertion should be given entirely by the $\alpha_{\Phi}$ and $L_5$ terms. Furthermore, estimates of
both $\alpha_{\Phi}$ and $L_5$ can be obtained from the hairpin analysis alone by first determining $\alpha_{\Phi}$ from the residue of
the single-pole term in the hairpin correlator (which gives $\alpha_{\Phi}\approx 0$), and then attributing the 
remaining mass dependence of the double-pole residue to the $L_5$ term.
Since the $L_5$ term is also responsible for the quark mass dependence of the axial-vector decay constant $f_A$, we have two
independent ways of estimating $L_5$ and hence a consistency check on the $Q\chi PT$ analysis of pseudoscalar correlators. 
Note also that, for
equal quark masses, quenched chiral logs are not present in either $f_A$ or $m_0^{eff}$. Thus linear quark mass dependence 
is expected and is in fact observed in our data \cite{BET}. 
Taking $\alpha_{\Phi}=0$, $Q\chi PT$ predicts \cite{Sharpe_hp} that the slope of $m_0^{eff}$ is equal in magnitude and opposite in
sign from that of $f_A$. Thus the product $m_0^{eff}\times f_A$ should be independent of quark mass,
and the negative slope of the mass insertion
may be inferred from the well-known positive slope of $f_A$. In this brief note, we examine the quantitative consistency 
of our data with this prediction. 

First we summarize the relevant Monte Carlo calculations from Ref. \cite{BET} and review the $Q\chi PT$ predictions for the 
mass dependence of $m_0^{eff}$ and $f_A$ \cite{Sharpe_hp}. The quark-connected (valence) and quark-disconnected (hairpin) 
pseudoscalar correlators are defined in momentum space by
\begin{equation}
\tilde{\Delta}_c(p) = \int d^4x e^{ip\cdot x}\langle\bar{q}_A\gamma_5 q_B(x)\;\tilde{q}_B\gamma_5 q_A(0)\rangle 
\end{equation}
and
\begin{equation}
\tilde{\Delta}_h(p) = \int d^4x e^{ip\cdot x}\langle\bar{q}_A\gamma_5 q_A(x)\;\tilde{q}_B\gamma_5 q_B(0)\rangle
\end{equation}
Here $q_A$ and $q_B$ are two quark fields with distinct flavor indices. For the results considered here, we take the two quarks
to have degenerate mass. Clover improved Wilson-Dirac fermions are used, with $C_{sw}=1.57$ for the $\beta=5.7$ configurations
and $C_{sw}=1.50$ for $\beta=5.9$. Quark propagators are calculated in the modified quenched approximation (MQA) by locating and shifting
real eigenmodes, as described in \cite{MQA}. Fitting the long-range time dependence of the Fourier transformed 
$\vec{p}=0$ valence correlator in the usual way, we extract the pion mass and pseudoscalar decay constant
from the location and residue of the propagator pole,
\begin{equation}
\tilde{\Delta}_c(p) = -\frac{f_P^2}{p^2+m_{\pi}^2} +  \ldots
\end{equation}

\begin{table}
\centering
\caption{Values for $m_{\pi}, m_0^{eff}$, and $f_A$ for $\beta=5.7$. Values are in GeV, using the rho mass
scale $a^{-1}=1.12$ GeV. Values of $f_A$ include a perturbative renormalization factor of $Z_A=0.845$.
Values of $m_0^{eff}$ should be multiplied by a flavor factor of $\sqrt{3}$ for comparison with the
physical $\eta'$ mass.}
\vspace*{0.5cm}
\label{tab:hairpin1}
\begin{tabular}{|c|c|c|c|}
\hline
$\kappa$ & $m_{\pi}$ & $m_0^{eff}$ &
$f_A$  \\
\hline
.1410  &  .566(2) & .314(11) & .173(2)  \\
.1415  &  .504(3) & .329(11) & .167(2)  \\
.1420  &  .432(3) & .345(11) & .160(2)  \\
.1423  &  .383(4) & .354(11) & .156(3)  \\
.1425  &  .344(4) & .360(11) & .154(3)  \\
.1427  &  .299(5) & .361(12) & .152(4)  \\
.1428  &  .274(6) & .361(12) & .152(5)  \\
\hline
\end{tabular}
\end{table}

\begin{table}
\centering
\caption{Values for $m_{\pi}, m_0^{eff}$, and $f_A$ for $\beta=5.9$. Here $a^{-1} = 1.64$ GeV,
and $Z_A = 0.865$.}
\vspace*{0.5cm}
\label{tab:hairpin2}
\begin{tabular}{|c|c|c|c|}
\hline
$\kappa$ & $m_{\pi}$ & $m_0^{eff}$ &
$f_A$  \\
\hline
.1382 & .674(5) & .318(8) & .172(1) \\
.1385 & .620(5) & .325(8) & .166(1) \\
.1388 & .563(7) & .333(8) & .160(2) \\
.1391 & .499(7) & .343(8) & .154(2) \\
.1394 & .428(8) & .356(8) & .147(2) \\
.1397 & .334(8) & .371(10) & .139(3)\\
\hline
\end{tabular}
\end{table}

\begin{figure}
\psfig{figure=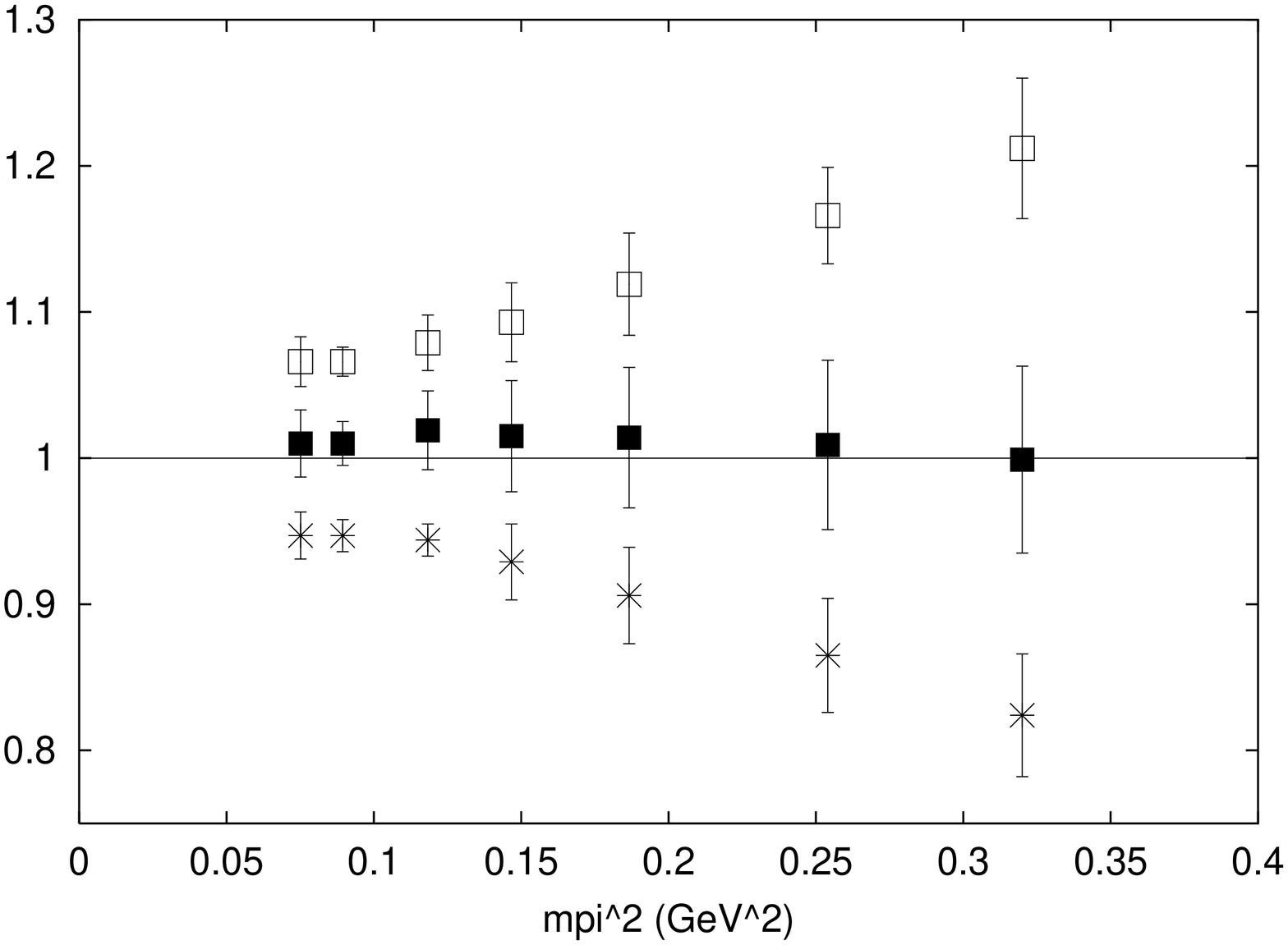,angle=270,width=1.0\hsize}
\psfig{figure=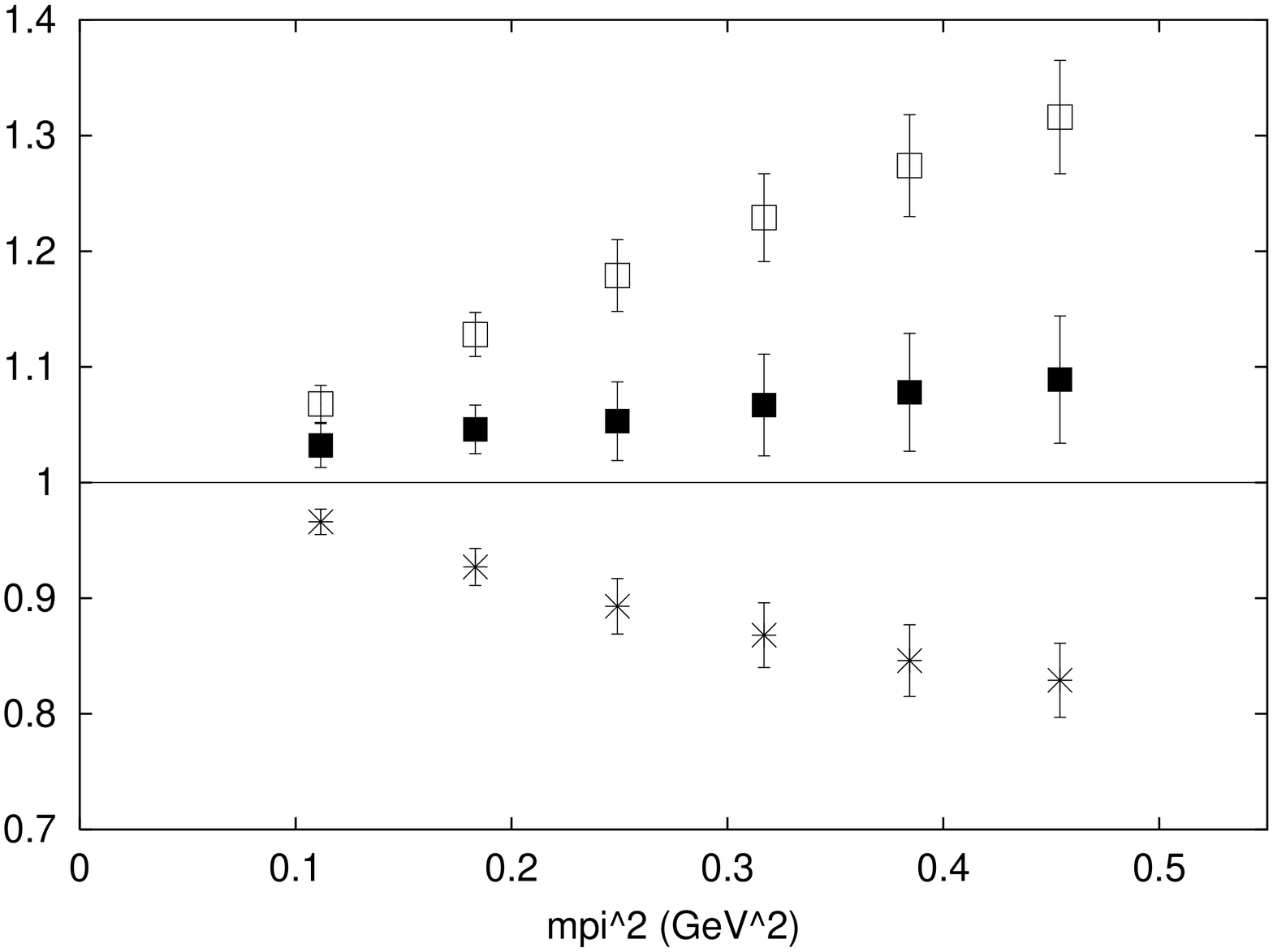,angle=270,width=1.0\hsize}
\caption[]{The ratios $R_f$ (open squares), $R_m$ ($\times$'s), and the product $R_f R_m$ (solid squares) for $\beta=5.7$ (upper plot)
and $\beta=5.9$ (lower plot). }
\label{fig:slopes}
\end{figure}

\vspace*{0.5cm}
The values obtained for $m_{\pi}$ are listed in Tables I and II, along with the values for the axial-vector decay 
constant $f_A$. The complete analysis is discussed in \cite{BET}.
We do not list the values of $f_P$ since they will not be used in the present
analysis, except to amputate the corresponding factors in the hairpin correlator. With pion masses fixed from the valence
correlator analysis, the hairpin correlators are fit to a combination of double- and single-pole terms,
\begin{equation}
\tilde{\Delta}_h(p) = (m_0^{eff})^2\frac{f_P^2}{(p^2+m_{\pi}^2)^2} +
\alpha_{\Phi}\frac{f_P^2}{p^2+m_{\pi}^2} + \ldots
\end{equation}
The values obtained for $m_0^{eff}$ are listed in Tables I and II. From the $Q\chi PT$ analysis in Ref. \cite{Sharpe_hp},
it is found that the double-pole residue is given, to first order in $m_{\pi}^2$, in terms of 
the Lagrangian parameters $m_0$, $\alpha_{\Phi}$, $L_5$, and $f$, by
\begin{equation}
(m_0^{eff})^2 = m_0^2\left(1-8L_5\frac{m_{\pi}^2}{f^2}\right) - \alpha_{\Phi}m_{\pi}^2
\end{equation}
As discussed in \cite{BET} the single pole coefficient
$\alpha_{\Phi}$ is small and consistent with zero. In the subsequent analysis, we will assume $\alpha_{\Phi}=0$.
With this simplification, $Q\chi PT$ predicts  that the chiral behavior of $m_0^{eff}$ is related to
that of $f_A$ by the requirement that their product is independent  of quark mass to first order, i.e.
\begin{equation}
\label{eq:prod}
m_0^{eff}\times f_A = {\rm const.} + O(m_{\pi}^4)
\end{equation}
To directly test this prediction, we would like not only to see that the slope of the product is consistent with zero,
but also that a significant slope in each of the two quantities separately is largely cancelled out in the product.
This is conveniently illustrated by plotting the ratio of each quantity to it's chirally extrapolated value. Define
the quantities
\begin{eqnarray}
\label{eq:ratio_f}
R_f(m_{\pi}^2) &=& \frac{f_A(m_{\pi}^2)}{f_A(0)}, \\
R_m(m_{\pi}^2) &=& \frac{m_0^{eff}(m_{\pi}^2)}{m_0^{eff}(0)}.  \label{eq:ratio_m}
\end{eqnarray}

In Figure 1 we have plotted the quantities $R_f$, $R_m$, and the product $R_f\times R_m$ as a function of
$m_{\pi}^2$. The errors that are plotted were computed by a jackknife analysis, with chiral extrapolation
carried out on each subensemble to determine the denominators in (\ref{eq:ratio_f}) and (\ref{eq:ratio_m}). 
The plots in Fig. 1 make it quite clear that the slopes of $f_A$ and $m_0^{eff}$ approximately cancel in the
product. For the data at $\beta=5.7$ the cancellation is quite accurate and the slope of the product is 
very close to zero. For $\beta=5.9$, the product appears to 
have a slight positive slope. A jackknife analysis directly on the slope of the product gives
\begin{equation}
\label{eq:prod2}
R_f  R_m = 1 + \epsilon m_{\pi}^2
\end{equation}
with $\epsilon = -0.051 \pm 0.16$ for $\beta=5.7$ and $\epsilon = 0.20\pm 0.14$ for $\beta=5.9$
to be compared with the individual slopes of about $\pm 0.60$ for $R_f$ and $R_m$ separately. 
The difference of this slope from zero in the $\beta = 5.9$ ensemble is of marginal significance and may be purely
a statistical fluctuation. It should be noted, however, that the numerical value of the clover coefficient $C_{sw}$  
for $\beta<6.0$ is not very well determined by theoretical considerations. It may be that our choice of $C_{sw}=1.50$
for $\beta=5.9$ leaves some residual $O(a)$ lattice spacing effects. The sensitivity of the slope in (\ref{eq:prod2})
to choice of clover coefficient has not been explored.

As emphasized by Sharpe \cite{Sharpe_hp}, the quark-mass dependence of the hairpin vertex provides a new way of 
determining the quenched Leutwyler parameter $L_5$.
The extraction of the Leutwyler parameter $L_5$ from the slope of $f_A$ was discussed in \cite{BET}. By a similar
analysis, an estimate of $L_5$ may be obtained from the slope of $m_0^{eff}$. Although the difference between
these two determinations is not very significant statistically, the values obtained from the $m_0^{eff}$ slope
are somewhat smaller than those obtained from $f_A$, particularly for the 5.9 data. From the jackknife subensemble
analysis, we find that the slopes of $f_A$ and $m_0^{eff}$ are {\it positively} correlated, i.e., since the slopes
are of opposite sign, the magnitude of the two slopes (and hence the corresponding 
estimates of $L_5$)fluctuate in opposite directions. This suggests
that we compare the slope of $f_A$ directly with the slope if $1/m_0^{eff}$.
To this end, we analyze the slope of $f_A$, $\sqrt{f_A/m_0^{eff}}$ and 
$1/m_0^{eff}$ in the form $A + Bm_{\pi}^2$.
The Leutwyler parameter is then given by 
\begin{equation}
L_5 = (1/8) \frac{B}{A} f_A(0)^2 = (1/4) \frac{B}{A} f^2 
\end{equation}
where the physical value of $f$ is $93$ MeV.

For $\beta = 5.7$, we find very consistent values,
\begin{eqnarray}
L_5 (f_A) &=& 1.66(21) \times 10^{-3} \\
L_5 (\sqrt{f_A/m_0^{eff}}) &=& 1.63(28) \times 10^{-3} \\
L_5 (1/m_0^{eff}) &=& 1.69(36) \times 10^{-3} 
\end{eqnarray}
were the errors have been separately determined from a jackknife analysis.

For $\beta = 5.9$, the results are not as consistent, reflecting the small residual slope
found for the product, $R_f R_m$, 
\begin{eqnarray}
L_5 (f_A) &=& 1.51(12) \times 10^{-3} \\
L_5 (\sqrt{f_A/m_0^{eff}}) &=& 1.27(14) \times 10^{-3} \\
L_5 (1/m_0^{eff}) &=& 1.09(15) \times 10^{-3}.
\end{eqnarray}
Here the value determined from $1/m_o^{eff}$ is somewhat smaller than the
result obtain from $f_A$ alone and is closer in magnitude to the $L_5$
values found at larger beta by the Alpha collaboration \cite{Alpha}.  We have
discussed other aspects of this comparison in a previous paper \cite{BET}.

We conclude that our analysis of the hairpin insertion vertex at $\beta =
5.7$ and $\beta = 5.9$ gives a striking confirmation of Sharpe's analysis
\cite{Sharpe_hp} of the chiral Lagrangian parameters and the relation between the
chiral slopes of $f_A$ and $1/m_o^{eff}$.

The work of W. Bardeen and E. Eichten was performed
at the Fermi National Accelerator Laboratory, which is
operated by University Research Association,
Inc., under contract DE-AC02-76CHO3000.
The work of H. Thacker was supported in part by the
Department of Energy under grant DE-FG02-97ER41027.

\end{document}